\documentclass[a4paper,11pt]{article}
\pdfoutput=1 
\usepackage{jheppub} 
\usepackage[]{epsfig}
\usepackage{amsfonts}
\usepackage{amsmath,bm}
\usepackage{subfigure}
\usepackage{graphicx}
\usepackage{graphics}
\usepackage{bbold}
\usepackage{slashed}

\def\ba{\begin{eqnarray}}
\def\ea{\end{eqnarray}}
\def\be{\begin{equation}}
\def\ee{\end{equation}}

\title{ Towards holographic flat bands }

\author[a,b]{Nicol\'as Grandi}\author[c,d]{Vladimir Juri\v ci\' c}
\author[a]{Ignacio Salazar Landea}
\author[e]{Rodrigo Soto-Garrido}

\affiliation[a]{Instituto de F\'\i sica de La Plata - CONICET, C.C. 67, 1900 La Plata, Argentina}
\affiliation[b]{Departamento de F\'\i sica - UNLP, Calle 49 y 115 s/n, 1900 La Plata, Argentina}

\affiliation[c]{Nordita, KTH Royal Institute of Technology and Stockholm University, Roslagstullsbacken 23, 10691 Stockholm, Sweden}
\affiliation[d]{Departamento de F\'isica, Universidad T\'ecnica Federico Santa Mar\'ia, Casilla 110, Valpara\'iso, Chile.}
\affiliation[e]{Facultad de F\'isica, Pontificia Universidad Cat\'olica de Chile, Vicu\~{n}a Mackenna 4860, Santiago, Chile}

\emailAdd{grandi@fisica.unlp.edu.ar}
\emailAdd{juricic@gmail.com}
\emailAdd{peznacho@gmail.com}
\emailAdd{rodrigo.sotog@gmail.com}

\abstract{Motivated by the phenomenology in the condensed-matter flat-band Dirac systems, we here construct a holographic model that imprints the symmetry breaking pattern of a rather simple Dirac fermion model at zero chemical potential.In the bulk we explicitly include the backreaction to the corresponding Lifshitz geometry and compute the dynamical critical exponent. Most importantly, we find that such a geometry is unstable towards a nematic phase, exhibiting an anomalous Hall effect and featuring a Drude-like shift of its spectral weight. Our findings should motivate further studies of the quantum phases emerging from such holographic models.
}

\begin{document}
\maketitle

%%%%%%%%%%%%%%%%%%%%%%%%%%%%%%%%%%%%%%%%%

\section{Introduction}

{Over the last decade, considerable research efforts were focused on the use of holography as a tool to illuminate  the properties of strongly correlated condensed matter systems. Understanding strange metallic,  superconducting and other phases in the phase diagram of high-T$_c$ cuprates is probably one of the most outstanding challenges. A summary of the large amount of effort done in this direction can be found in Refs.~\cite{Zaanen-2015,hartnoll2018}. In this context,  the recently uncovered twisted bilayer graphene (TBG)~\cite{cao2018correlated,cao2018unconventional}, which features a very rich phase diagram, extensively studied both theoretically ~\cite{venderbos2018,rademaker2018,guinea2018,guo2018,kang2018,koshino2018,kennes2018,liu2018,ochi2018,wu2018,yuan2018,kang2019,seo2019,ahn2019,gonzalez2019,hejazi2019a,hejazi2019b,huang2019,lian2019,you2019,po2019,roy2019,tarnopolsky2019,wu2019,bultinck2020,fernandes2020,abouelkomsan2020,bernevig2020a,bernevig2020b,bernevig2020c,cea2020,christos2020,huang2020,kang2020,konig2020,ledwith2020,padhi2020,repellin2020a,repellin2020b,soejima2020,vafek2020,wu2020,xie2020a,xie2020b,Mxie2020,fernandes2021,potasz2021,liu2021} and experimentally~\cite{yankowitz2019,lu2019,sharpe2019,kerelsky2019,polshyn2019,choi2019,jiang2019,serlin2020,cao2020,chen2020,liu2020,lu2020,park2020,saito2020,stepanov2020untying,Swu2020,choi2020,nuckolls2020,wong2020,saito2021,choi2021},  stemming from the enhanced interaction effects in the flat bands forming in the system, may be of importance. This rich landscape of possible phases and  phase diagrams together with seeming  similarities with the ones of high-T$_c$ superconductors~\cite{andrei2020graphene}, motivate us to formulate and study a holographic model that might encode a possible pattern of symmetry breaking in TBG.}

In this context, we emphasize that a first realization of flat bands in a holographic model was proposed in Ref.~\cite{Laia:2011zn} . In this approach,  Lorentz symmetry breaking, a necessary ingredient for constructing flat bands, is realized through the boundary conditions imposed on a Dirac
spinor field living in an $AdS$ background geometry. As such, this model describes a flat band, but it does not include the backreaction to the geometry. 
%and the violation of Lorentz invariance can be thought of as a subleading effect in powers of $N$.

We here propose a different approach to address the symmetry breaking, which explicitly includes the backreaction, and is based on the method employed to construct the holographic duals to Weyl semimetals \cite{landsteiner2015,Landsteiner:2015pdh}, multi-Weyl semimetals \cite{Dantas:2019rgp}, and PT-symmetric non-hermitean systems \cite{Arean:2019pom}. We start by constructing a toy model for flat bands in terms of free Dirac fermions at the neutrality point (zero chemical potential). We then rewrite this model by invoking  an emergent global symmetry, which is preserved in the UV, and use simple relevant sources that break such symmetry to produce the flat bands in the IR. In the next step, we use the standard holographic dictionary and promote the global symmetries to gauge symmetries in the bulk. We subsequently construct geometries with boundary conditions given by the relevant deformations studied in the free model. Such a  holographic construction, as we show here, at intermediate scales between UV and IR, yields Lorentz-symmetry breaking Lifshitz geometry, as expected for a flat band. We emphasize that the backreaction is explicitly included  to obtain the dynamic exponent $z$ characterizing the Lifshitz geometry. Most importantly, we find that such a geometry below a critical temperature is unstable towards a nematic phase exhibiting  an anomalous Hall effect. Finally, the off-diagonal conductivity features a Drude-like shift of its spectral weight, which, in a condensed matter model, may be associated with the formation of Fermi pockets in a nematic phase~\cite{calderon2020correlated}.

\section{Free fermions}
Let us first discuss the flat bands within a  free Dirac fermion model.  The corresponding  Dirac Hamiltonian describing a free fermion at neutrality in $2+1$ spacetime dimensions reads
\begin{equation}
    H_D= -\gamma^t(\gamma^x p_x+\gamma^y p_y)\,,
\end{equation}
where $\gamma^\mu= (\sigma_3,-i\sigma_2,i\sigma_1)$ are the $2+1$ Dirac gamma matrices. The eigenstates of this Hamiltonian are characterized by the relativistic dispersion relation $\omega=\pm \sqrt{p_x^2+p_y^2}$.

We now couple two copies of free Dirac fermion, analogously to the approach used in Ref. \cite{Dantas:2019rgp}, to construct a non-relativistic system as a relevant deformation of this relativistic one \cite{Pena-Benitez:2018dar,Hoyos:2020ywe}. We emphasize here that this approach is analogous to the construction of the effective low-energy Hamiltonian for bilayer graphene by coupling two single layers each featuring linearly dispersing Dirac fermions. A first possibility is to write
\begin{equation}
    H_D'= H_D \otimes \mathbb{1}_{\mbox{\scriptsize$2\!\times\!2$\normalsize}}+ i\,{m_*} \left(\gamma^x \otimes \sigma_2 -\gamma^y \otimes \sigma_1  \right)\,.
\label{H}
\end{equation}
This Hamiltonian is still quadratic and the spectrum reads
\begin{equation}
    \omega=\pm m_* \pm \sqrt{p_x^2+p_y^2+{m_*}^2}
\end{equation}
with gapped conduction and valence bands corresponding to both positive or both negative signs respectively, while otherwise the valence and conduction bands cross at zero energy. In the latter case, the low-energy excitations feature a quadratic dispersion relation for $p_x^2+p_y^2\ll m_*$
\begin{equation}
    \omega\approx \pm \frac 1{2m_*}\left(p_x^2+p_y^2\right)
\end{equation}
which still preserves rotational invariance. 

Although Gaussian, this model features quite an interesting hierarchy of scales, which may be interpreted in terms of a renormalization group flow. In the deep UV we start with the two decoupled relativistic Dirac fermions. Now, we turn on the coupling, which is a relevant perturbation, and at the end of the RG flow, at a scale much smaller than  the coupling between the two species ($m_*$), one of them is gapped out while the remaining one becomes non-relativistic, giving a fixed point with dynamic critical exponent  $z=2$. Interestingly, the IR fixed point has a larger $z$ than the original one, seemingly counterintuitive from a naive power-counting point of view. The price to pay is that close to this new fixed point with a larger dynamic exponent, the quartic electron interactions, by virtue of having a scaling dimension $z-2$ after taking into account quantum fluctuations, become more relevant than close to the Dirac vacuum, and destabilize it. 

The wave equation resulting from \eqref{H} reads $i\,\partial_t\Psi=H'_D\Psi$ in terms of the fermionic pair $\Psi=(\psi_1,\psi_2)$ where $\psi_1$ and $\psi_2$ are two-component Dirac fields. It can be obtained from the action
\begin{eqnarray}
    S&=&
    %\int d^3x\, \left(\Psi^\dagger \,i \partial_t  \Psi-\Psi^\dagger H'_D\Psi\right)=
    \int d^3x\, i\bar\Psi\left(\gamma^\mu\partial_\mu  -i\,{m_*}  \left(\gamma^x \otimes \sigma_1  +\gamma^y \otimes \sigma_2\right)\right)\Psi 
%\nonumber\\    &=&
=S_{\sf free}-i\int d^3x\, \bar\Psi \slashed{W}\Psi.
    \label{S}
\end{eqnarray}
Here $S_{\sf free}$ is the free Dirac action for the pair of spinors, which exhibits a $U(2)$ global symmetry as well as spatial rotational invariance. On the other hand, the deformation corresponds to coupling the pair to a constant non-Abelian vector field $W=m_*(\sigma^{1}dx+\sigma^{2}dy)$, which explicitly breaks the $U(2)$ symmetry down to $U(1)$.  Notice that even if spatial rotational invariance were broken from the outset, any $SO(2)$ rotation acting on the vector $W$ could be undone by a $U(2)$ transformation in the direction of $\sigma_3$, implying that there is a preserved rotational invariance in the system. We will therefore construct a holographic 
%dual to a flat band 
model
by taking  a bulk theory  with a $U(2)$ gauge symmetry, whose boundary conditions  explicitly break it to a residual $U(1)$, while preserving a combination of rotational and $U(1)_3$ invariances.  

It is important to stress that we are not building the holographic theory dual to the model \eqref{H}, but to a strongly coupled theory with the same symmetry breaking pattern. In other words, we are replacing $S_{\sf free}$ in  Eq.~\eqref{S} by a strongly coupled action invariant under spatial rotations and a global $U(2)$ symmetry.

\section{The holographic theory}

As explained in the previous section, in addition to the standard AdS gravitational sector, the bulk theory must contain a gauge $U(2)=U(1)\times SU(2)$ invariance. The minimal action with such requirements reads
%
% \begin{eqnarray}
% S= - \int d^4x\sqrt{-g}\left(R-2\Lambda
% +
% {\color{red}
% \frac{1}{4}} F\wedge {^{\!\star}  F} +
% \mathrm{Tr}\left(
% G\wedge{^{\!\star} G}\right)
% \right)\,,\label{eq:HoloAct}
% \end{eqnarray}
%
\begin{eqnarray}
S= S_\text{grav}
-\frac{1}{4}\int
\left[ F\wedge {^{\!\star}  F} +
\mathrm{Tr}\left(
G\wedge{^{\!\star} G}\right)
\right]\,,\label{eq:HoloAct}
\end{eqnarray}
with 
$$S_\text{grav}=\int d^4x\sqrt{-g}\left(R-2\Lambda\right)$$
and where ${F}$ represents the $U(1)$ gauge field strength while $G$ is the strength of $SU(2)$ gauge field. They are defined from the corresponding gauge  fields ${A}$ and $B=B^a\,\sigma_a/2$ with $a=1,2,3$ in the standard way ${F}=d A$ and $G=d{B} - i(q/2) {B}\wedge B$.  

To construct a holographic theory we look for asymptotically AdS solutions in the above theory. 
Since our aim is to describe a flat band, we have to turn on a relevant deformation that breaks the boundary $U(2)$ group down to $U(1)$. This translates into the boundary condition
\begin{equation}\label{eq:boundcond}
 B\xrightarrow[r\to 0]{}{m_*}\left(\sigma_1\mathrm dx +\sigma_2\mathrm dy\right) =W\, .
\end{equation}
%
%{\color{red} in what follows $m_*=\Delta/2$.}
This constraint breaks the rotational symmetry in the $xy$ plane as well as the $U(2)$ gauge symmetry, but it preserves a combination of the two \cite{Gubser:2008zu}, yielding what  we will refer to as  ``rotational symmetry'' in the following. An ansatz for the background fields consistent with the sources we turned on at the boundary reads
\begin{eqnarray}~\label{eq:ansatz}
 B =  \frac12\left(Q_{1}(r) \sigma_1\,+Q_{2}(r) \sigma_2\,\right) dx +  \frac12\left( Q_{1}(r)\sigma_2+Q_{2}(r)\sigma_1 \right)dy\,,
 \,\,\, {A}=0\,.
\end{eqnarray}
Here we recover the minimal ansatz when setting $Q_2=0$. Nevertheless, from Ref.~\cite{Juricic:2020sgg} we know that spontaneous symmetry breaking of rotational invariance might happen at low temperatures. Hence we extend the minimal ansatz to include the so called $xy$-nematic phase when  $Q_2\neq0$. We chose that combination because of its stability in the 5d model \cite{Juricic:2020sgg} {and it corresponds to a nematic phase since it breaks the rotations $SO(2)$ down to a discrete $C_4$. }

For the metric, on the other hand, we write
\begin{equation}
\mathrm d s^2 = \frac{1}{r^2}\left(-N(r)f(r)\mathrm dt^2 + \frac{dr^2}{f(r)} + dx^2 + dy^2+2 h(r)dx\, dy \right).\label{eq:metrica}
\end{equation} 
For the rotational invariant phase with $Q_2=0$, it is consistent to set $h=0$.
The requirement of having AdS asymptotics then translates into the boundary conditions 
\begin{eqnarray}
 f\xrightarrow[r\to0]{} 1\,,\qquad\qquad N\xrightarrow[r\to0]{} 1\,,\qquad\qquad 
 h\xrightarrow[r\to0]{} 0\,.
 \label{eq:asymp}
\end{eqnarray}
Plugging the ansatz given by Eqs.~\eqref{eq:ansatz}-\eqref{eq:metrica} into the equations of motion obtained from \eqref{eq:HoloAct}, we find the set of coupled differential equations of the form

\begin{eqnarray}
\label{eq:san}
-\frac{4r^2hh'}{N}\left(Nf\right)'&=&-8(3+rf')(1-h^2) +  q^2 r^4 \left(Q_1^2 - Q_2^2\right)^2  + \\ 
&& 2f \left(12(1-h^2) - 3 r^2 h'^2 + 
    8 r^4 h Q_1' Q_2' + 
    4 r^4 (Q_1'^2 + Q_2'^2) (1-2h^2) \right)
\nonumber\\ 
-\frac{ (1-h^2)4r \left(Nf^2\right)'}{fN}&=&24 (1-h^2)- q^2 r^4 (Q_1^2 - Q_2^2)^2 + \\ 
&&  4f (-6(1- h^2) +r^2h'^2- 
    4r^4h Q_1' Q_2' + 
    2r^4h^2 (Q_1'^2 + Q_2'^2)-2rhh')
\nonumber\\
 \left(Q_1' \sqrt{N}f\right)'&=&\frac{ \sqrt{N}}{4(h^2-1)}\left(q^2 (Q_1 - h Q_2) (Q_2^2 - Q_1^2) -
 4 f h' Q_2'\right) \\
\left(Q_2' \sqrt{N}f\right)'&=&\frac{ \sqrt{N}}{4(h^2-1)}\left(q^2 (Q_2 - h Q_1) (Q_1^2 - Q_2^2) -
 4 f h'Q_1'\right) \\
\left(\frac{h'f \sqrt N}{r^2}\right)'&=&2f  \sqrt{N}\left(-2 Q_1'Q_2'+h\left( Q_1'^2+Q_2'^2 +\frac{h'^2}{2r^2(h^2-1)}\right)\right). 
\label{eq:guche}
\end{eqnarray}

In the following, we explore solutions to these equations of motion subject to the boundary conditions given by Eqs.~\eqref{eq:boundcond} and \eqref{eq:asymp}. The IR boundary conditions depend on whether we are analyzing the zero temperature or the finite temperature cases, as also discussed below. 

\section{Zero temperature solutions}

In this section we will focus on domain wall solutions that interpolate between an IR and a UV fixed point at zero temperature. We first identify an exact isotropic Lifshitz solution that can be interpreted as a candidate IR geometry from which we can shoot into the boundary conditions \eqref{eq:boundcond}-\eqref{eq:asymp} in the UV. Nevertheless, for a sufficiently large backreaction parameter $q$ this exact IR solution turns out to be unstable under anisotropic perturbations. We then change the IR solution by a pure AdS one, which is, in turn, stable under anisotropic perturbations, and study numerically its deformation into a solution satisfying the boundary conditions given by Eqs.~\eqref{eq:boundcond}-\eqref{eq:asymp} in the UV. The resulting numerically obtained profiles describe a nematic phase, in which rotational symmetry is spontaneously broken.

\subsection{Lifshitz solution}

\begin{figure}[t!]
    \centering
    \includegraphics[width=0.483\textwidth]{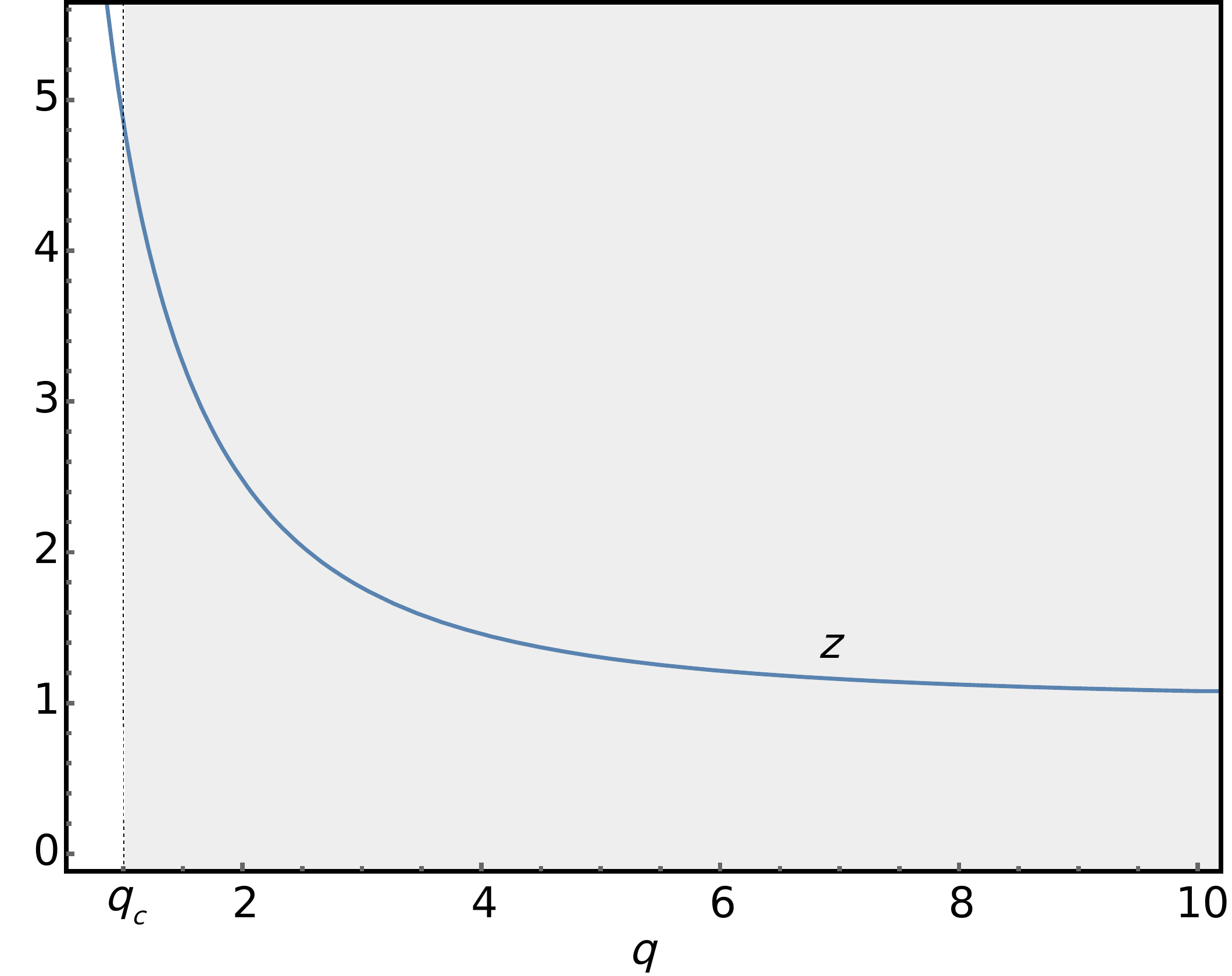}\hfill\includegraphics[width=0.504\textwidth]{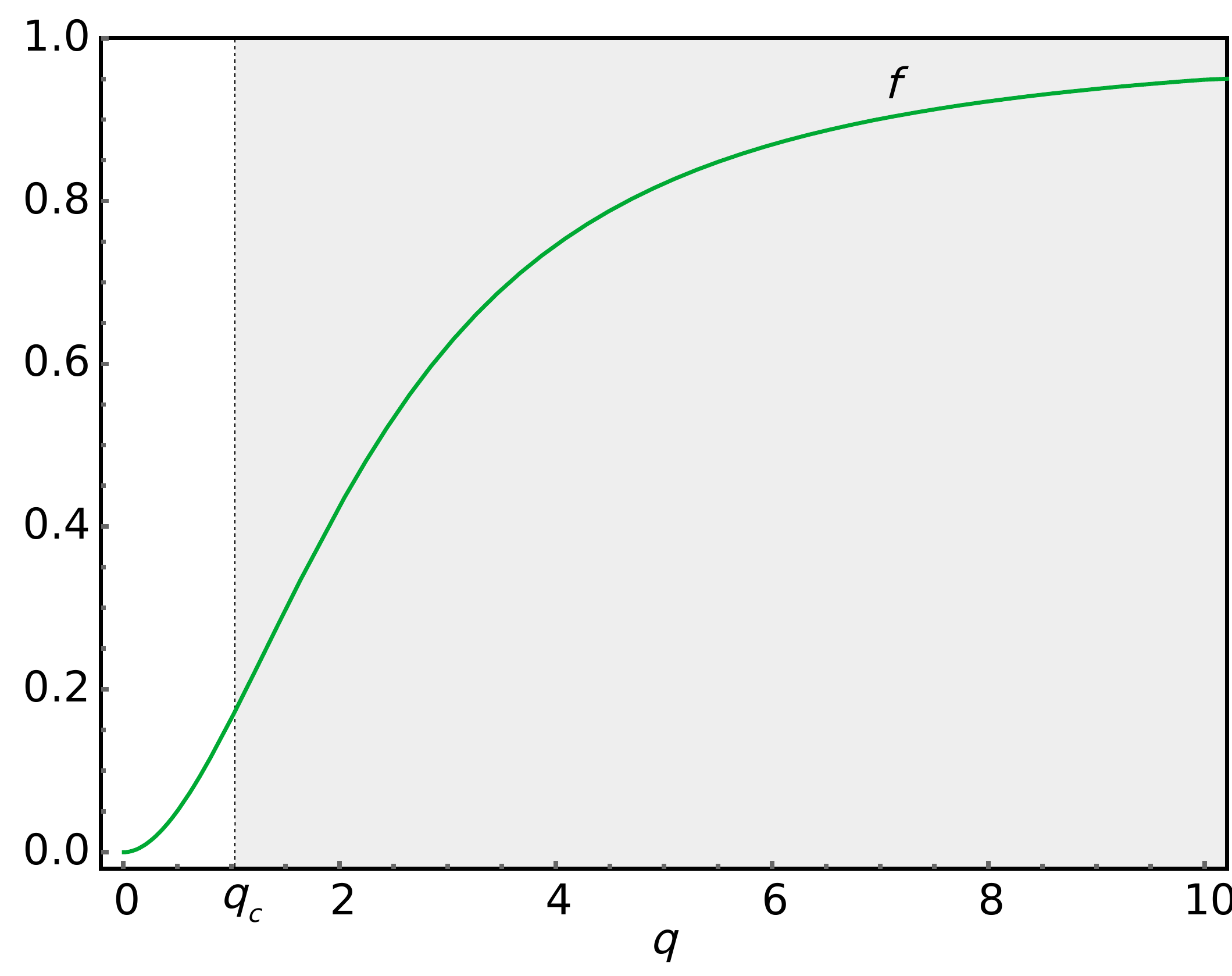}
    \caption{ Left: Dynamical critical exponent $z$ as a function of the backreaction parameter $q$, as given by Eq.~\eqref{eq:zq}. Right: The number of the degrees of freedom $f$, given by Eq.~\eqref{eq:f1}, as a function of the backreaction parameter $q$. In both plots, the gray region corresponds to values  $q>q_c$ for which the Lifshitz solution is unstable.}
    \label{fig:zvsq}
\end{figure}

Let us start by searching for the fully rotationally symmetric  solutions. In other words, we replace $Q_2=h=0$ in the Eqs.~ \eqref{eq:san}-\eqref{eq:guche}. This simplifies the equations considerably,
as can be checked by a direct substitution, and they are of the form
{
\begin{eqnarray}
 \label{Lsolsan}
 \label{eq:f1}
f&=&%\displaystyle
 \frac{24-(1-z)^2q^2}{8 \left(2+z\right)}\,,
 \\
 N&=&\displaystyle\frac{ N_L }{r^{2(z-1)}}\,,
\\
 Q_1&=&\displaystyle\frac{\sqrt{z-1}}{r}\,,
 \label{Lsolgughe}
\end{eqnarray}
}
where the dynamic exponent $z$ solves the following equation
\begin{equation}\label{eq:zq}
    q^2(z^3+z^2)+(q^2-24)z-3(q^2+8)=0\,.
\end{equation}
% }\
This cubic equation for the values of the parameter $q\gg1$ yields $z\approx1$, as shown in Fig.~\ref{fig:zvsq}. In other words, in this limit the Yang-Mills fields decouple from the Einstein equations, leaving the metric to be pure AdS, as expected. 
As we lower the value of $q$ the IR metric becomes a Lifshitz geometry with \emph{increasing} dynamic exponent $z$. For $q$ very small, we find $z\approx2\sqrt 6/q$, as can be seen in Fig.~\ref{fig:zvsq}. 
We also mention that the effective number of degrees of freedom in the IR theory, $f(q)$, as given by Eq.~\eqref{Lsolsan}, asymptotically approaches one as $q\rightarrow\infty$, see right panel in Fig.~\ref{fig:zvsq}.

The above exact isotropic Lifshitz solution can only be taken as a candidate for the IR geometry, since it does not satisfy the correct boundary conditions in the UV. To construct the full isotropic Lifshitz background, we must add an irrelevant perturbation that takes the solution away from the one described by Eqs.~\eqref{Lsolsan}-\eqref{Lsolgughe}, and triggers an RG flow towards the original relativistic UV fixed point, given by Eqs. ~\eqref{eq:boundcond}-\eqref{eq:asymp}.

However, as we will show next, such an isotropic Lifshitz background is generically unstable towards a nematic phase \cite{Cremonini:2014pca}, thus explicitly breaking the rotational symmetry and featuring $z=1$, for large enough $q$. Hence the full geometry we found is strictly speaking unstable, and as such can only approximate an intermediate scaling regime  \cite{Donos:2017ljs,Donos:2017sba,Arias:2018mqn}.

%\subsubsection{Nematic perturbation}

\begin{figure}[t!]
    \centering
    \includegraphics[width=0.6\textwidth]{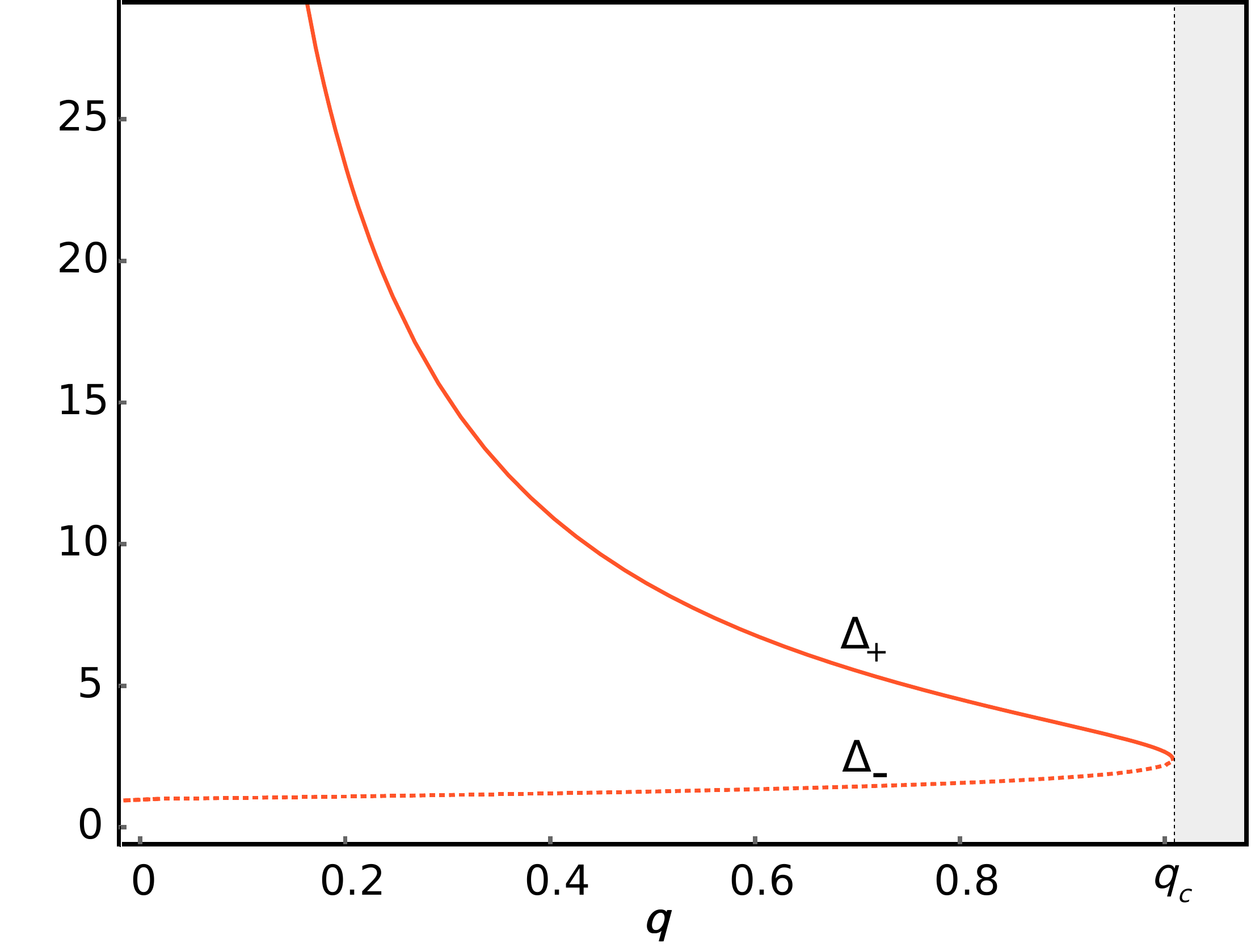}
    \caption{ Exponents $\Delta_\pm$ as a function of the backreaction parameter $q$,  given by Eq.~\eqref{eq:DeltaPlusMinus}. The upper (solid) line is to the $\Delta_+$ curve, while the lower (dotted) one corresponds to the $\Delta_-$. As before, the gray region corresponds to values  $q>q_c$ for which the Lifshitz solution is unstable.}
    \label{fig:nemins}
\end{figure}

To study the stability of the isotropic Lifshitz background, we expand the equations of motion around its exact IR form  \eqref{Lsolsan}-\eqref{Lsolgughe}  to linear order. In particular, the one corresponding to $Q_2$ can be solved by 
\begin{equation}
    Q_2\approx c_+ r^{\Delta_+}+c_-r^{\Delta_-}
\end{equation}
where $c_\pm$ are arbitrary constants, and the exponents $\Delta_\pm$ read
\begin{equation}\label{eq:DeltaPlusMinus}
    \Delta_\pm=\frac12\left(z\pm \sqrt{z^2-\frac{q^2}{f}\sqrt{z-1}\,}\right).
\end{equation}
These constants represent the effective conformal dimension of the dual operator in the deep IR. A sufficient condition indicating an instability is satisfied when such an effective dimension becomes complex, as it happens here for $q\gtrsim q_c\approx 1.00785$. As we will see, such an instability may still appear for $q \lesssim q_c$ at the nonlinear level. In Figure \ref{fig:nemins} we show $ \Delta_\pm$ as a function of $q$ for the range of $q$ where the solutions are real.

We have found so far that the isotropic Lifshitz background is generically unstable for large enough values of $q$. Later on, we will heat up the system and show how the form given by Eqs.~\eqref{Lsolsan}-\eqref{Lsolgughe}  plays an important role at intermediate temperature scales, leaving its imprint on the behavior of the entropy as a function of the temperature.

\subsection{Nematic domain walls}
\begin{figure}[t!]
    \centering
    \includegraphics[width=0.48\textwidth]{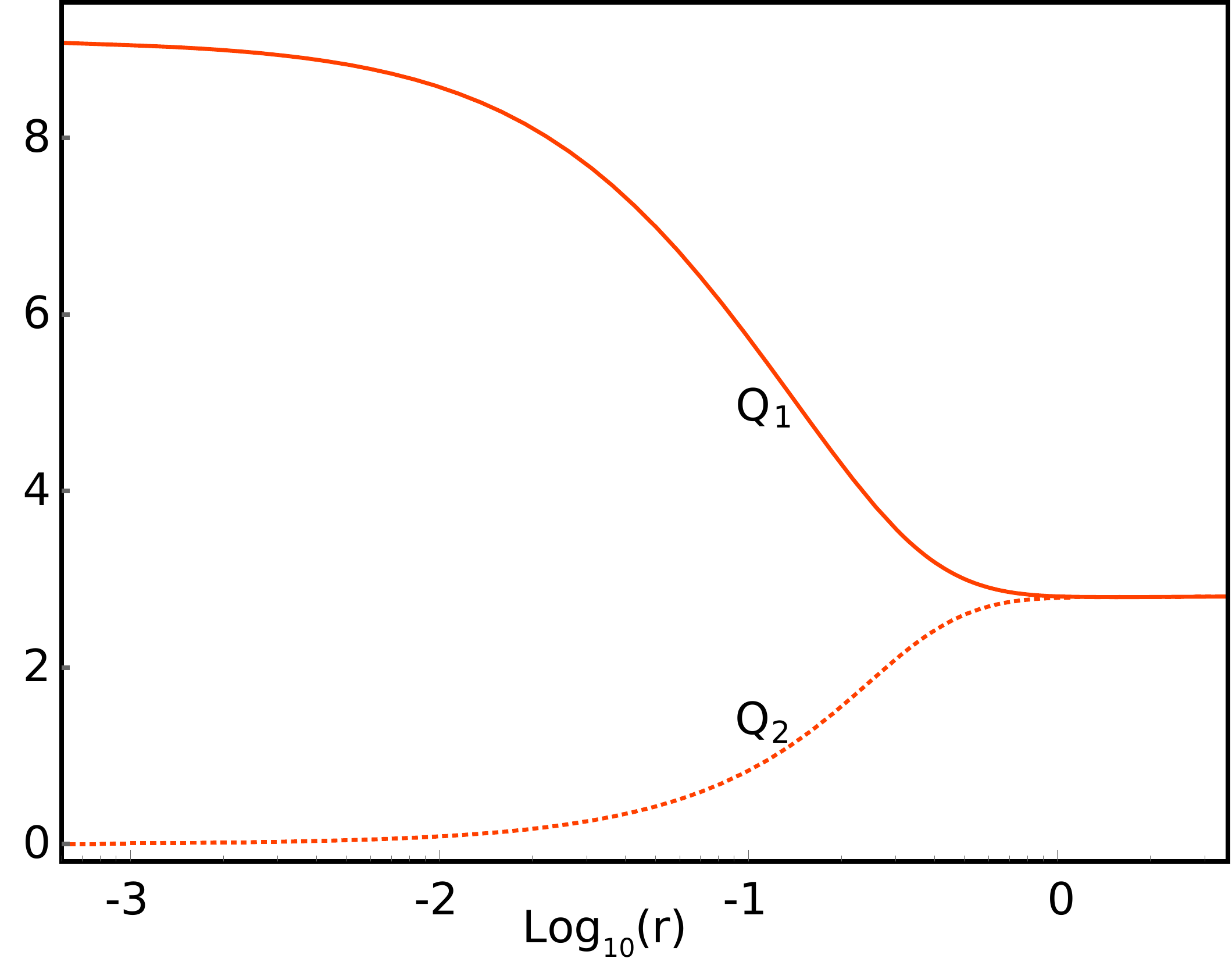}\hfill\includegraphics[width=0.48\textwidth]{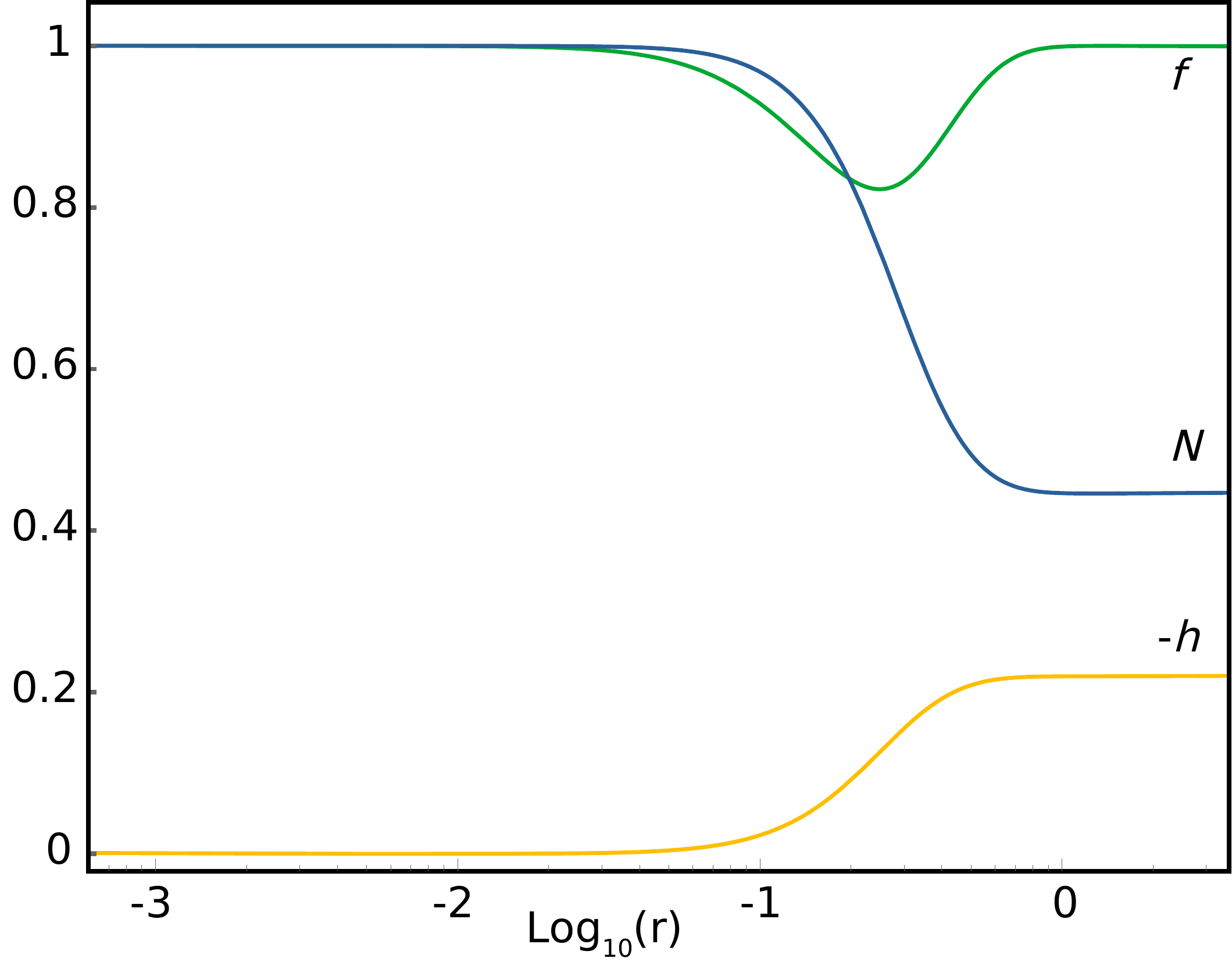}
    \caption{Radial profiles for a nematic domain wall at $q=2$. Left: $Q_1$ (upper line) and $Q_2$ (lower line). Right: The functions $f$ (upper line), $N$ (middle line) and $-h$ (lower line). These profiles are given by integrating from Eqs~\eqref{eq:f}-\eqref{eq:h}.}
    \label{fig:npro}
\end{figure}

Since for large enough $q$ the IR region of the isotropic Lifshitz solution becomes unstable, in order to find the stable background we need to replace such an IR form by a stable one. To this end, we choose constant IR values for all the functions $f=1, N=N_n, Q_1=Q_2=Q_n, h=h_n$.   This corresponds to pure AdS, since a constant $h$ can be removed by a change of variables,  and the same can be done with  constants $Q_1$ and $Q_2$ by applying a gauge transformation. In other words, the exact conformal invariance $z=1$ is recovered in the IR. 
This is no longer true as we extend the solution towards the UV
\begin{eqnarray}
 \label{eq:f}f&\approx& 1-\frac{2Q_p^2}pr^3 e^{-2pr}+\dots\,,\\
 \label{eq:Nn} N&\approx& N_n+\frac{2N_nQ_p^2}{p}r^3 e^{-2pr}+\dots\,,\\
 Q_1 &\approx & Q_n+\frac{Q_p}{p} e^{-pr}+\dots\,,\\
  Q_2& \approx & Q_n-\frac{Q_p}{p} e^{-pr}+\dots\,,\\
\label{eq:h}  h&\approx& h_n+{\frac{(1+h_n)^{3/2}Q_p^2}{p}r^2 e^{-2pr}}+\dots\,.
\end{eqnarray}
In this expansion $p={qQ_n}/\sqrt{1+h_n}$, and the non-trivial dependence in $r$ is just an expansion in powers of $r$ and $ \exp({-pr})$. 
Now we can integrate numerically away from the IR towards the desired UV using $Q_p,\,h_n,\,N_n$ as shooting parameters. We can shoot towards any numerical value for $m_*$ since the residual scaling symmetry allows us to reabsorb its numerical value. Hence, the value of $Q_n$ can be set to any non-zero numerical value, and we do not use it to shoot. An example of the profile for the field is shown in Figure \ref{fig:npro}. 

\begin{figure}[t!b]
    \centering
    \includegraphics[width=0.5\textwidth]{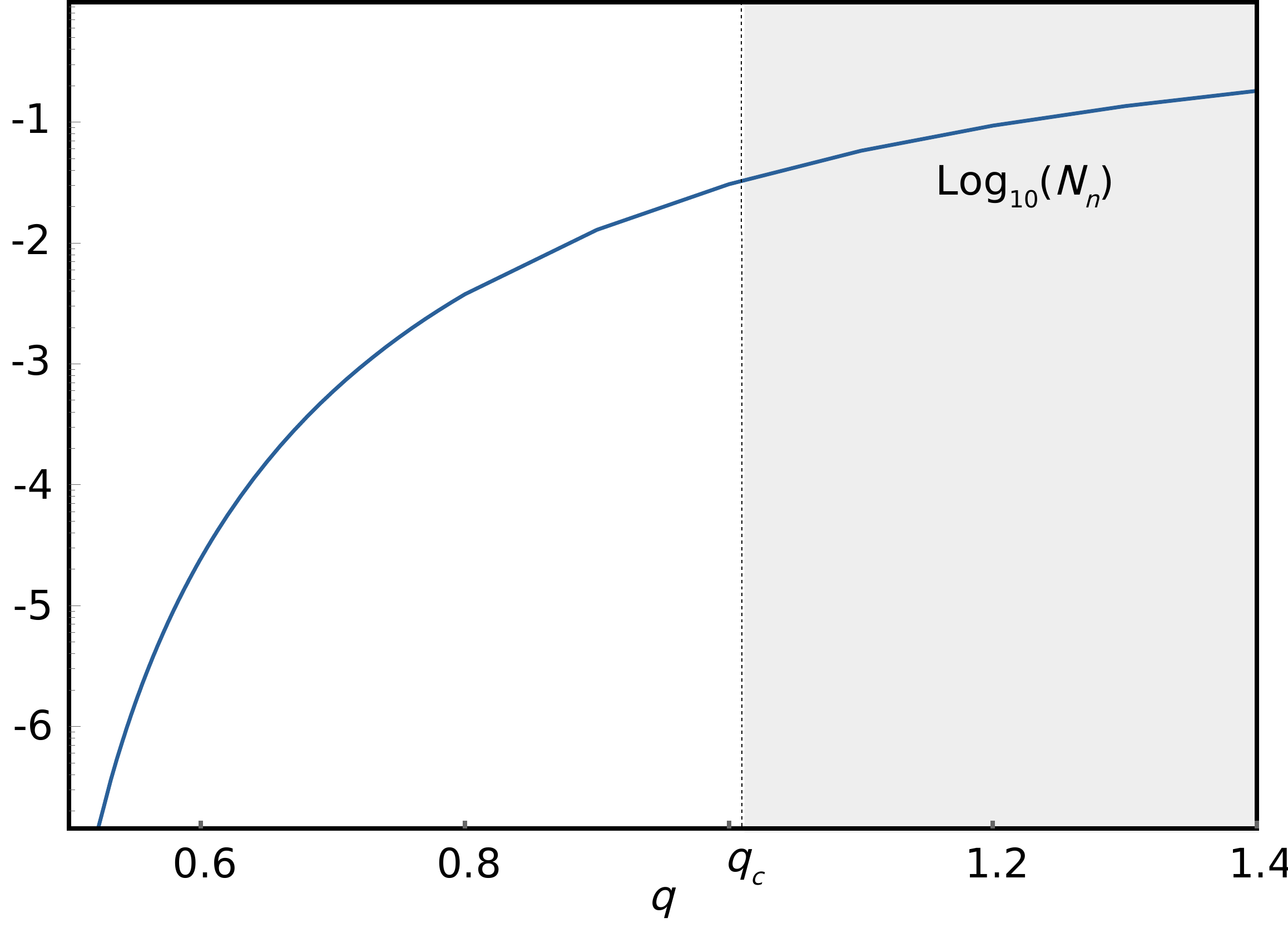}
    \caption{ The coefficient $N_n$ of equation (\ref{eq:Nn}) gets fixed after shooting towards the desired UV boundary conditions (\ref{eq:asymp}). We plot its value as a function of the backreaction parameter $q$ to see that solutions still exist below $q_c$, but the numerical value of $N_n$ becomes highly suppressed.  
    }
    \label{fig:nex}
\end{figure}

As the parameter $q$ is decreasing, we find solutions up to $q\geq 0.5362$, which is  below the critical value $q \approx 1.00785$ suggested by the linear IR behavior. We were not able to determine whether solutions beyond that critical value exist, since the numerical computations become too challenging in that regime. 
From Fig.~\ref{fig:nex} we may conclude that as $z$ increases and $q$ 
thus decreases,  the numerical computations become problematic.  This is so because 
the numerical value of $N_n$ becomes smaller and smaller, with the form that can be fitted by a function $\log N_n\approx 29 -23.3/q$. 

{Finally we would like to comment that the IR (\ref{eq:f}) corresponds to and AdS geometry with the same radius than the UV which represents a so called ``Boomerang RG flow'' \cite{Donos:2017sba}. Since our relevant deformation breaks Lorentz invariance it escapes the   hypothesis of holographic c-theorems \cite{Myers:2010xs,Myers:2010tj}. }

\section{Heating up: Finite-temperature solutions}

\begin{figure}[htb!]
     \centering
   \includegraphics[width=0.485\textwidth]{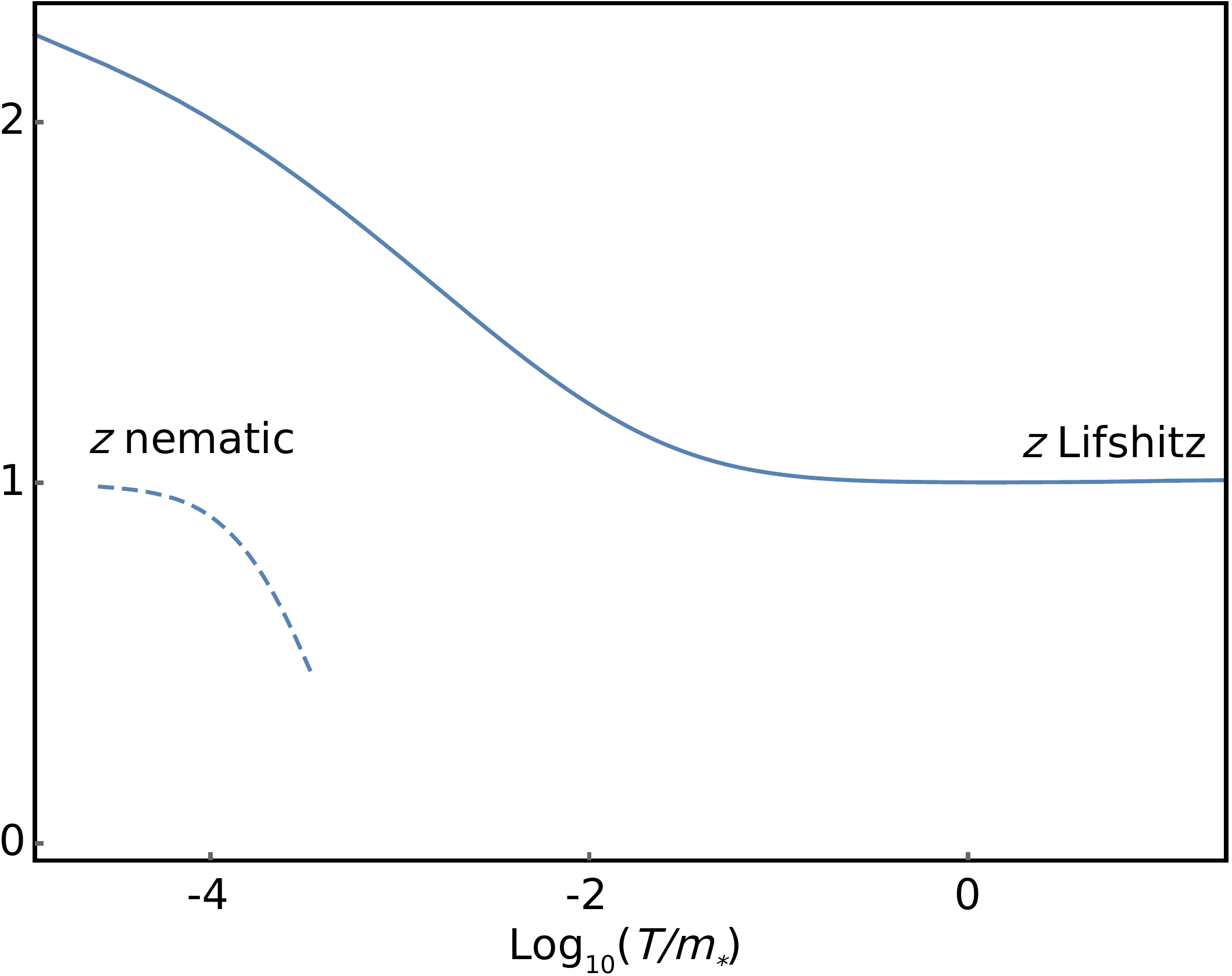}\hfill \includegraphics[width=0.485
    \textwidth]{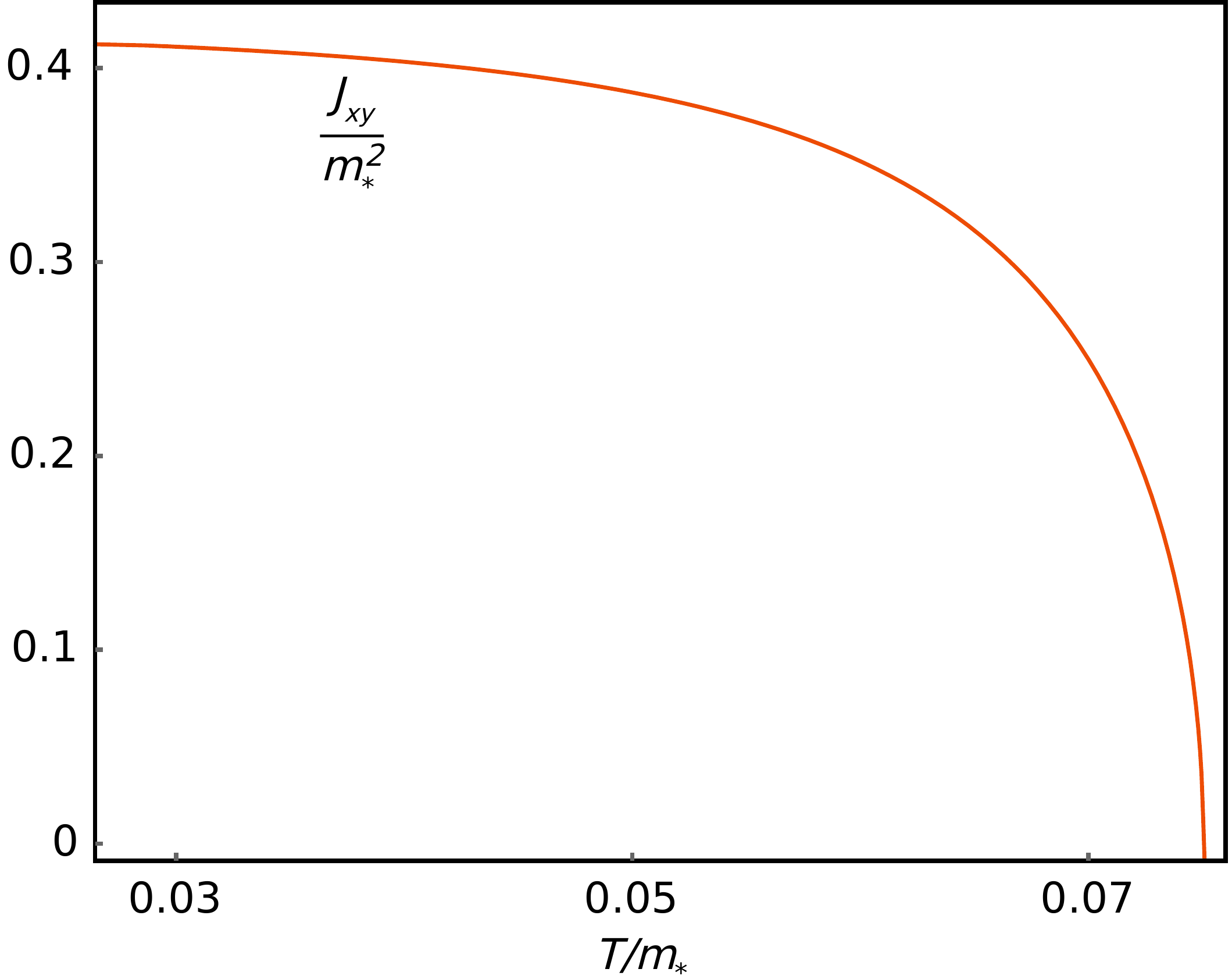}
    \caption{ Left: Dynamical critical exponent for the Lifshitz and nematic phases as a function of the temperature defined as $z\equiv{2S}/{TS'}$.
    The curves correspond to the $q=2$ solution. Right: Order parameter $J_{xy}$, defined as the subleading term in the near boundary expansion of $Q_2$, as a function of the temperature.}
    \label{fig:fT}
\end{figure}

Now we turn to study finite temperature solutions, characterized by the presence of a black hole horizon at some finite value of the radial coordinate $r=r_h$ where $f(r_h)=0$. The dual theory has a finite temperature which is determined by the near horizon expansion
\begin{eqnarray}
 f&\approx & \frac{4 \pi T}{\sqrt{N_h}}(r_h-r)+\dots\\
 N&\approx & N_h+\dots\\
 Q_1&\approx & Q_{1h}+\dots\\
 Q_2&\approx & Q_{2h} +\dots\\
 h&\approx & h_h+\dots.
\end{eqnarray}
 The temperature that satisfies the equations of motion close to the horizon is of the form
\begin{equation}
    T=\frac{\sqrt{N_h}\left(-12+12h_h^2+\frac12 q^2(Q_{1h}^2-Q_{2h}^2)r_h^4\right)}{16\pi(h_h^2-1)r_h}.
\end{equation}
In holography, it is then natural to fix $q$ and use $T/m_*$ as the dimensionless tuning parameter  to explore the phase diagram.
At high $T/m_*$ we only find rotational invariant solutions. 
From the scaling of the entropy with respect to the temperature $S\sim T^{2/z}$ we can read off the dynamical exponent, see Fig.~\ref{fig:fT}. %We show this scaling of the entropy as a function of the temperature in Figure (\ref{fig:fT}). 
This analysis then  confirms that the solutions interpolate from $z=1$ in the UV to $z\approx2.4$ (we set $q=2$ in the numerical computations) at low temperatures, in agreement with our previous findings at $T=0$.
For low enough temperatures, at certain $T=T_c$ the system undergoes a second order phase transition towards a nematic phase. In Fig. \ref{fig:fT} we show the order parameter $J_{xy}$, defined as the subleading term in the near boundary expansion of $Q_2$, as a function of the temperature. Furthermore, we checked from the on-shell action that the xy-nematic solution is preferred. The sub-leading coefficient in the $Q_1$ expansion will be non-zero for any value of $T/m_*$ as the operator has an explicit source turned on. 
In the low temperature regime, we observe a scaling of the entropy that is consistent with the dynamic exponent $z=1$ for the xy-nematic solutions.

\begin{figure}[htb!]
    \vspace{.4cm}
    \centering
    \includegraphics[width=0.58\textwidth]{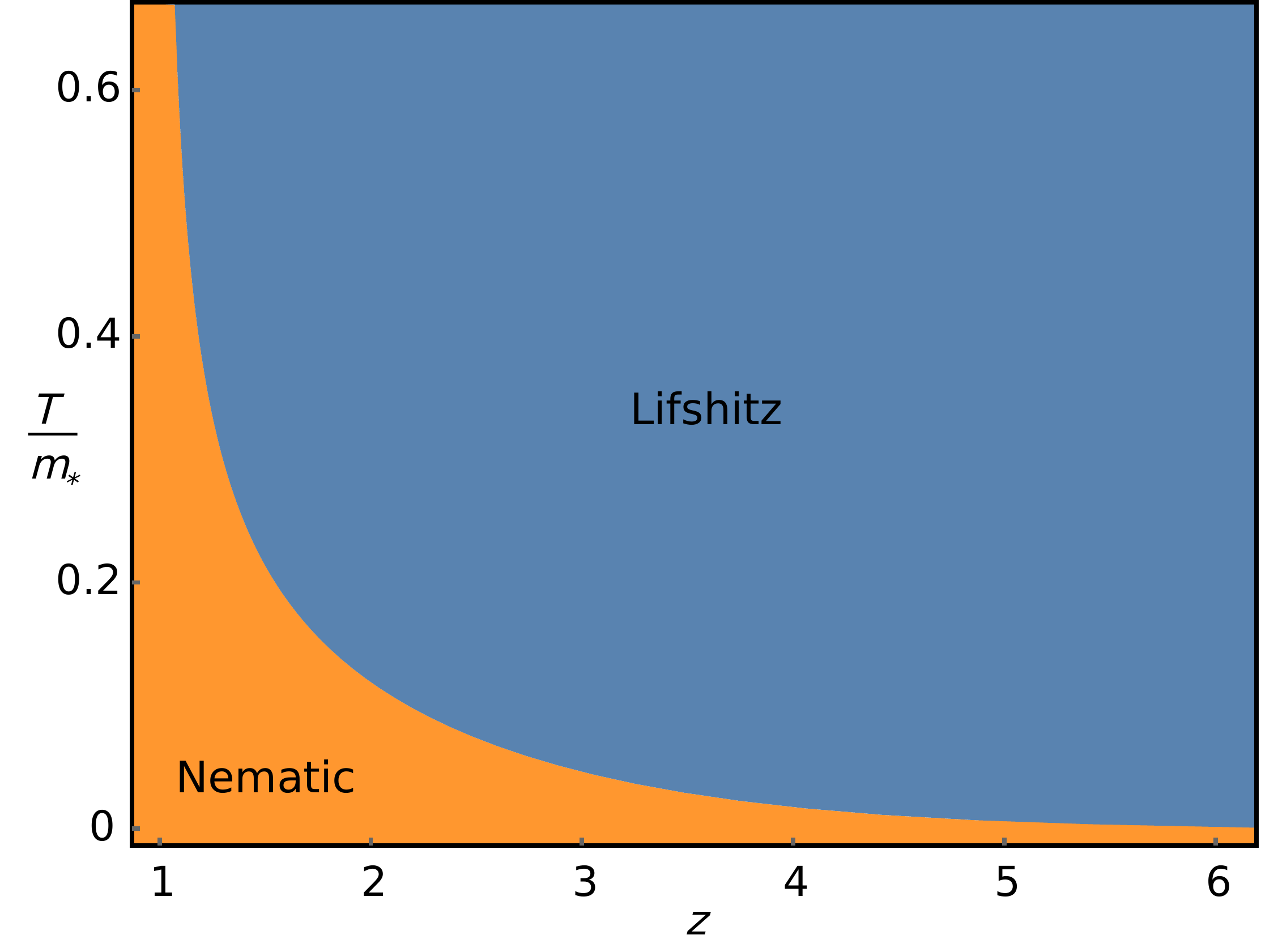}
    \caption{Phase diagram in the temperature versus dynamical critical exponent plane. We see that the critical temperature for the nematic phase decreases as the value of the critical exponent increases. { The horizontal axis corresponds to $z=z(q)$, defined as a solution to Eq. \eqref{eq:zq}}}
    \label{fig:pd2}
\end{figure}

We carried out the previous analysis at  fixed parameter $q$. On the other hand, as the parameter $q$ decreases, which implies an increasing the dynamical exponent $z$, the critical temperature gets smaller, making the nematic condensate less and less favored, as we show in Figure \ref{fig:pd2}. This behavior nicely fits the expectations form our $T=0$ analysis, but seems to 
contradict what could have been expected from a purely perturbative point of view. In general, one expects that as the density of states increases with the dynamical exponent, instabilities are enhanced at large $z$.
This behavior looks rather puzzling  but it may originate from the strongly coupled nature of the system, circumventing the expectations based on the weak coupling (perturbative) and single particle pictures. 

\subsection{Anomalous Hall effect}

\begin{figure}[t!]
    \centering
    \includegraphics[width=0.485\textwidth]{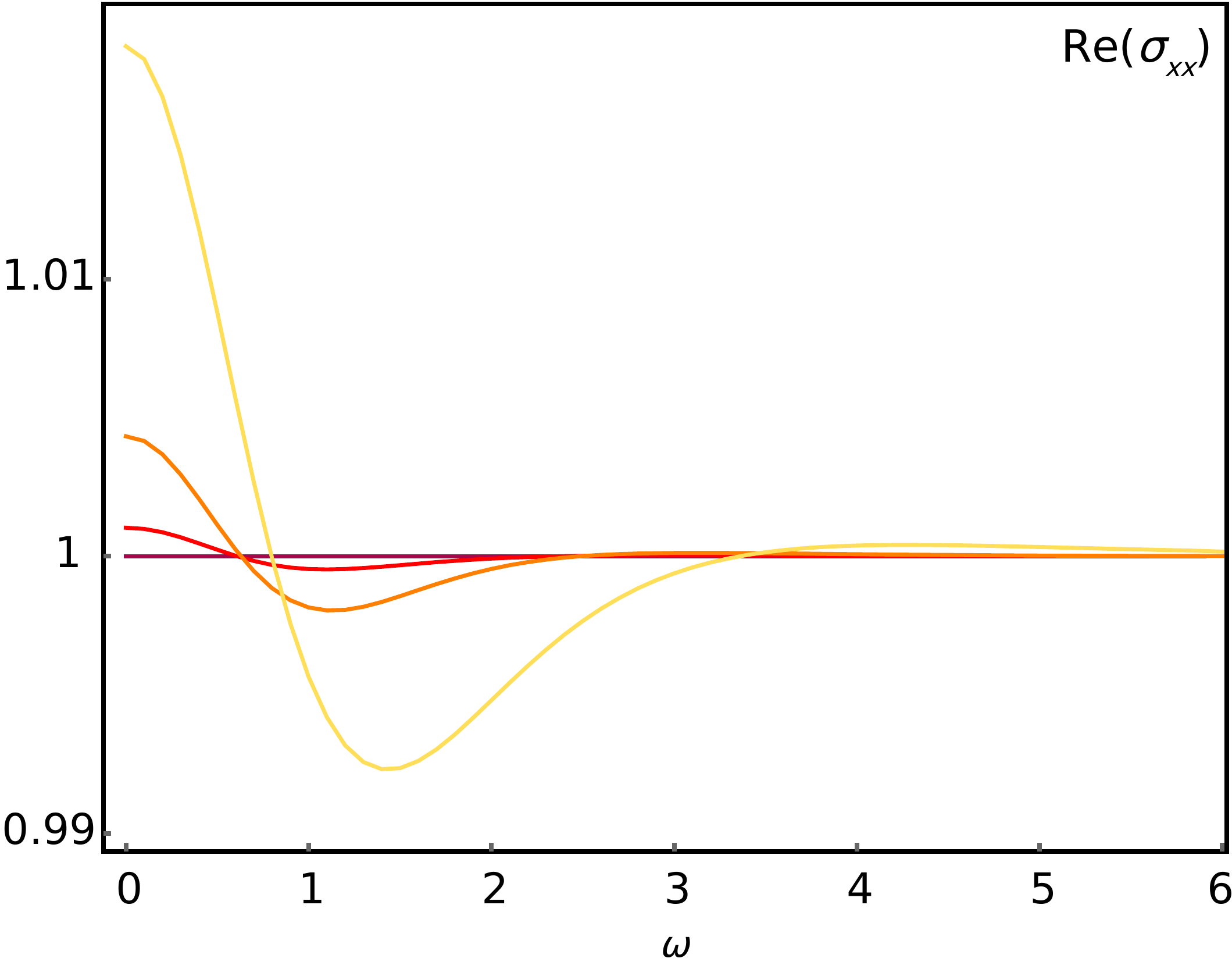}\hfill\includegraphics[width=0.485\textwidth]{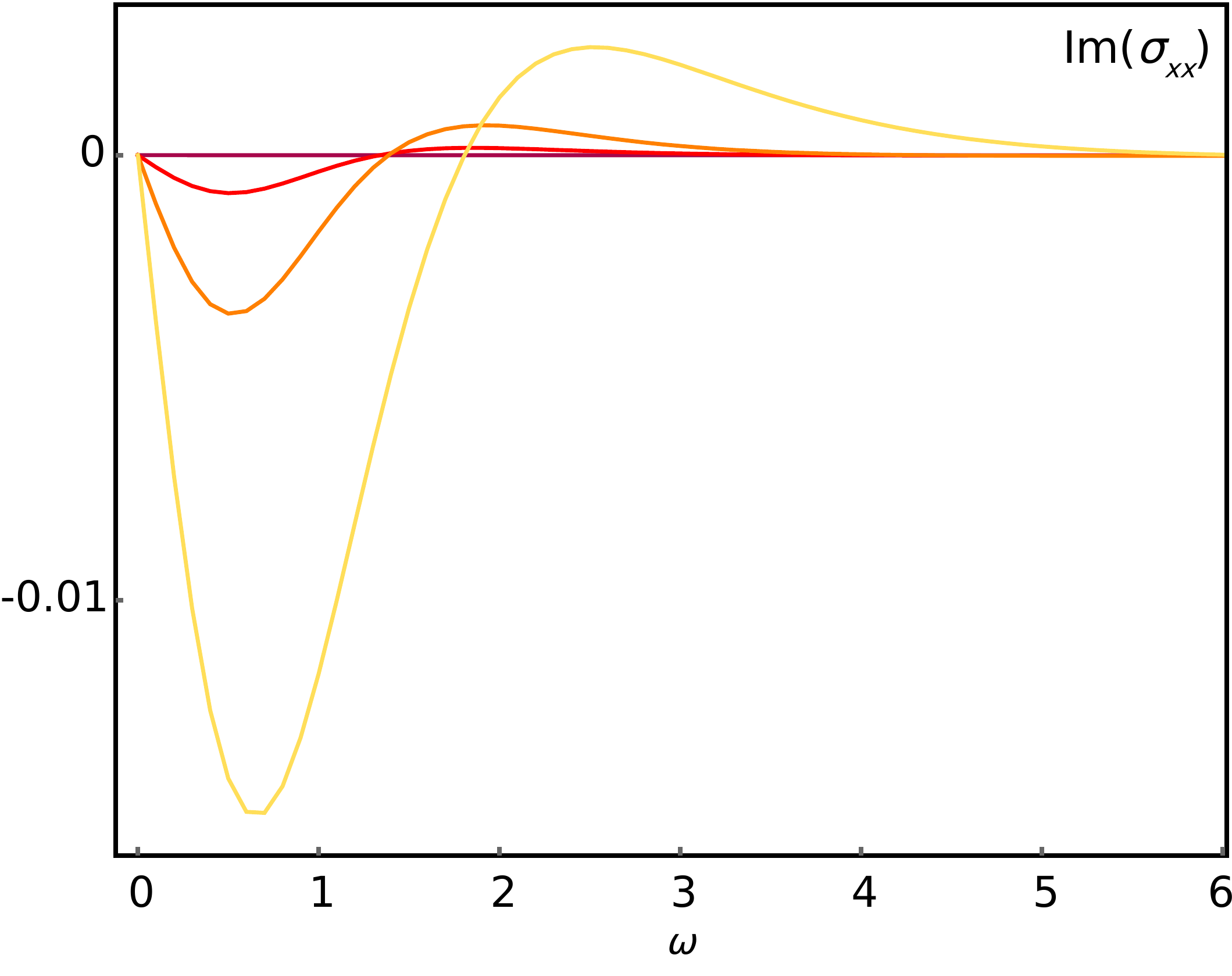}
\\ \vspace{.2cm}
    \centering
    \includegraphics[width=0.485\textwidth]{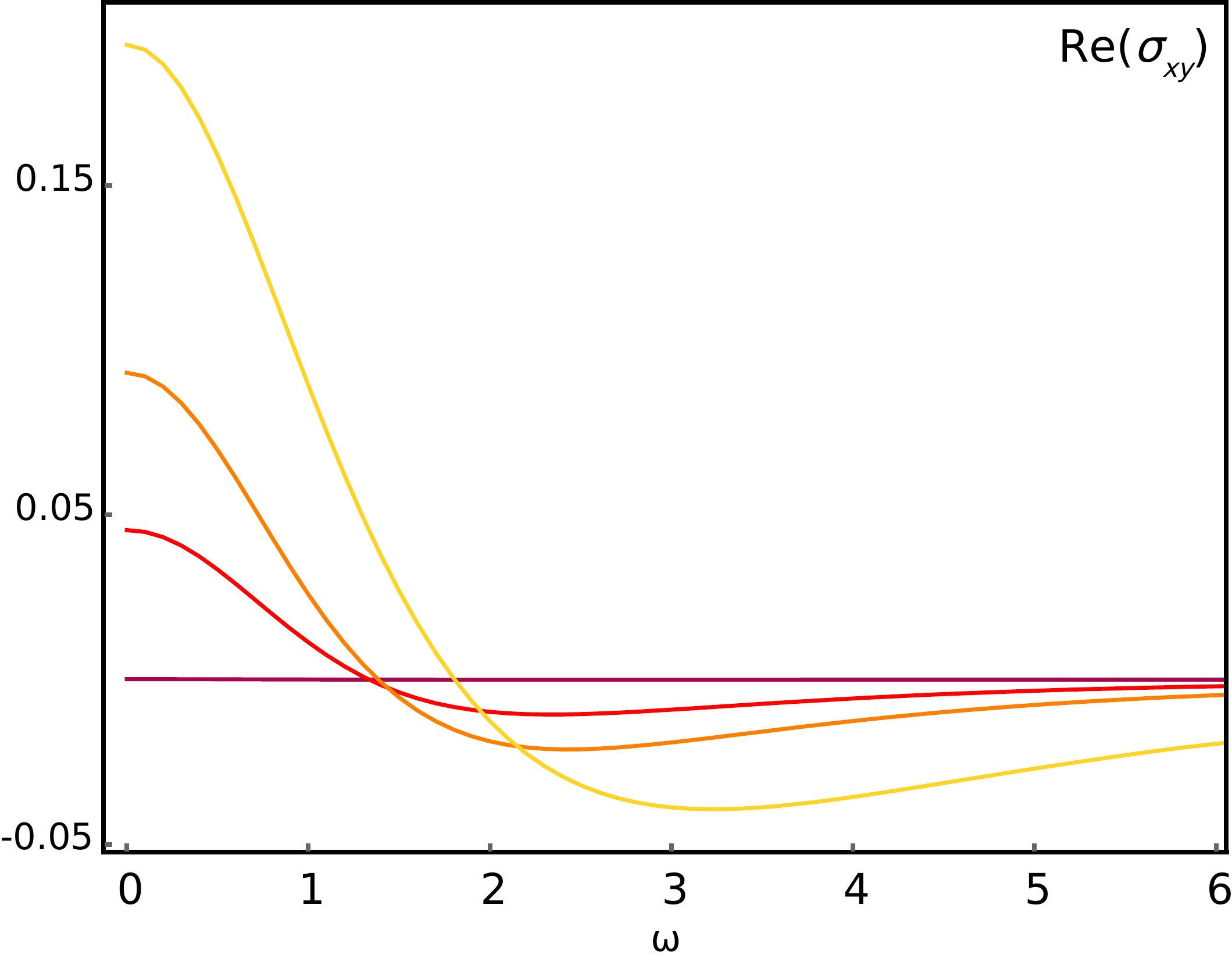}\hfill\includegraphics[width=0.485\textwidth]{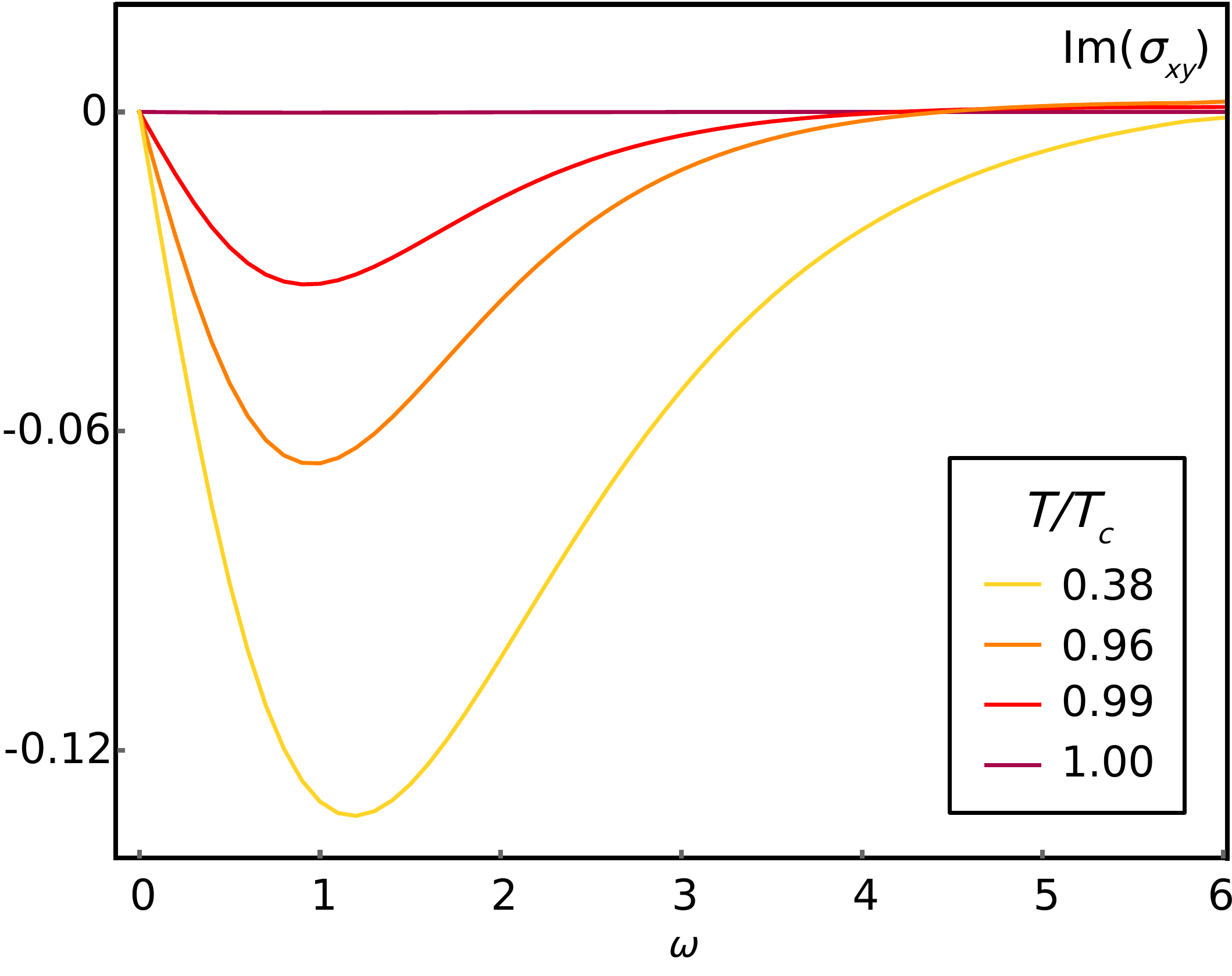}
    \caption{Top panels: real (left) and imaginary (right) parts  of the conductivity  $\sigma_{xx}$ as a function of the frequency for various values of $T/T_c$. Bottom panels: real (left) and imaginary (right) parts  of the Hall conductivity  $\sigma_{xy}$ as a function of the frequency for various values of $T/T_c$.}
    \label{fig:sxy}
\end{figure}

Now let us  turn to the study of the optical conductivities of the solutions constructed above. 
{ This calculation was carried out using standard methods as outlined in references \cite{Hartnoll:2007ai,Hartnoll:2009sz}}.
We consider a linear perturbation of the form 
\begin{equation}
    \delta A = e^{-i\omega t} \left(a_x(r)  dx+a_y(r) dy\right)\,,
    \end{equation}
% {\color{red} where $s_0$ is the $2\times2$ identity matrix}. 
giving the following equation for the $a_x$ component
\begin{eqnarray}
    a_x''= \left( -\frac{f'}{f}-\frac{N'}{2N}  \right)a_x'-\frac{\omega^2}{f^2N}a_x+\frac{h}{1-h^2}a_y'.
    \label{eqpert}
\end{eqnarray}
Since the $xy$-nematic phase is invariant under $x\to y$, $y\to x$, the equation of motion for $a_y$ is obtained from the above one by the exchange $a_y\rightarrow a_x$, $a_x\rightarrow a_y$. Notably, the xy-nematic background couples the fields $a_x$ and $a_y$ as is evident from the last term in Eq.~(\ref{eqpert}). In turn, this gives rise to a non-zero anomalous Hall conductivity.

Imposing infalling boundary conditions at the horizon
\begin{equation}
    a_i\approx a_{i(h)} (r-r_h)^{\frac{i\omega}{4\pi T}}+\dots
\end{equation}
we are left with two free coefficients at the horizon, $a_{x(h)}$ and $a_{y(h)}$, which we can use to get two independent solutions to the equations of motion. That will suffice to obtain the conductivity matrix $\bm{\sigma}$
\begin{equation}
    J_i=\sigma_{ij}E^j
\end{equation}
where we can read the currents $J_i$ from the subleading behavior for $a_i $ near the boundary, while the electric field is obtained from the leading behavior of $a_i$. We show the resulting conductivities in Fig.~ \ref{fig:sxy}.

 For large $\omega$ we find $\sigma_{xx}\to1$ and $\sigma_{xy}=0$, since the CFT governs the UV physics. At intermediate and low frequencies, on the other hand, both conductivities change from being featureless in the normal phase to oscillatory behavior as we dive deep into the nematic phase. Furthermore,  $\sigma_{xx}$ develops a Drude like peak, as we show in the top panels of Fig. \ref{fig:sxy} . A plausible explanation for this kind of shift in the spectral weight in the absence of momentum dissipation might be related to some additional effects, such as the formation of Fermi pockets in the nematic phase, as suggested for a condensed matter model~\cite{calderon2020correlated}.
 
\begin{figure}[htb!]
    \vspace{.5cm}
    \centering
    \includegraphics[width=0.485\textwidth]{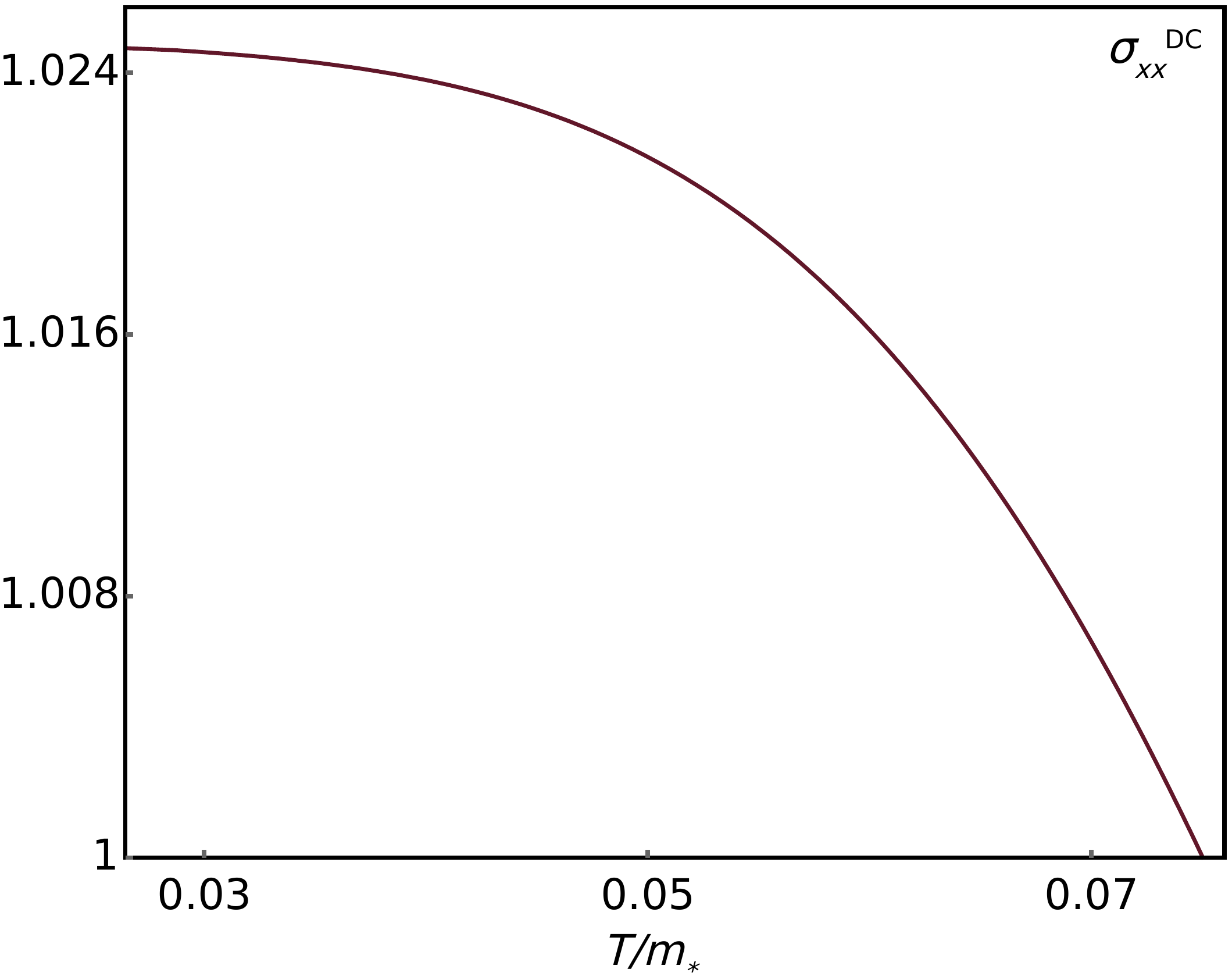}\hfill\includegraphics[width=0.485\textwidth]{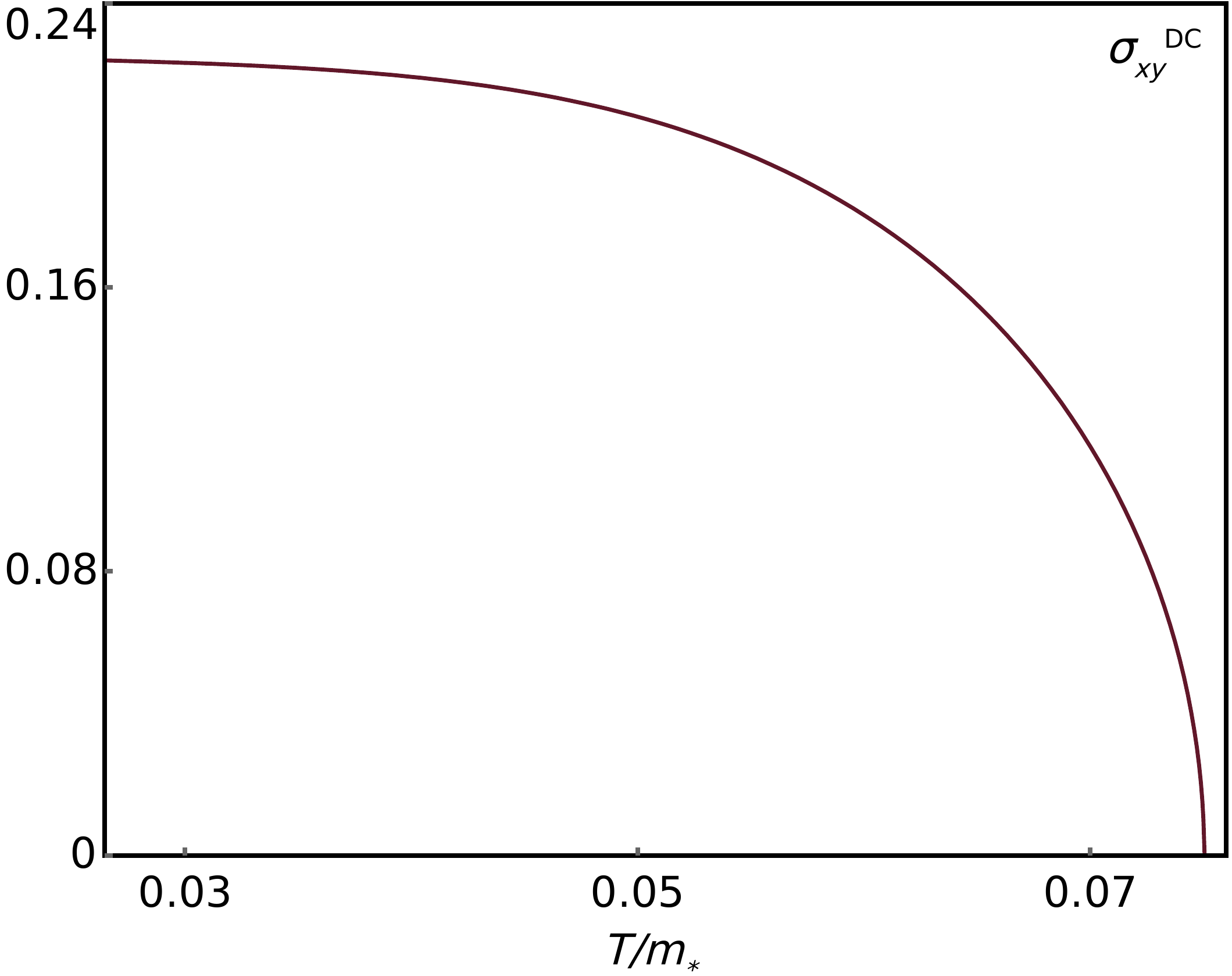}
    \caption{DC conductivities, defined as $\sigma_{ij}^{DC}\equiv\sigma_{ij}(\omega\to0)$, as a function of $T/m_*$.}
    \label{fig:sdc}
\end{figure}

 Interestingly, the anomalous Hall DC conductivity  $\sigma_{xy}^{DC}=\sigma_{xy}(\omega\to 0)$ has a non-zero real part which increases as the temperature is lowered and therefore the system is entering deeper into the nematic phase.  We observe that such anomalous conductivity is a good {indicator} for the nematic phase {\cite{landsteiner2015}}, see Fig.~\ref{fig:sdc}.

\section{Conclusions and outlook}

In summary, we here provided an explicit construction of flat bands within a holographic setup which explicitly takes into account the symmetries of the free Dirac fermions, and that includes the backreaction. As we showed, a nematic instability can emerge from the flat bands constructed in this manner,  which exhibits anomalous Hall conductance. Finally, the off-diagonal optical conductivity in the nematic phase features a Drude-like shift of the spectral weight.

The analysis of the flat bands introduced in this paper can be extended in several directions. First, one can add the chemical potential and study the instabilities in this setting. In particular, it would be interesting to see whether the nematic instability survives at a finite chemical potential. Furthermore,   adding  a charged operator under the global $U(1)$ symmetry in this setting would allow to study the interplay between the nematic phases and a holographic superconducting order \cite{Hartnoll:2008vx,Hartnoll:2008kx}, which we plan to study in a future.
Also, as we are claiming to have an approximately flat band at an intermediate scale, %and Fermi pockets at the nematic phase that dominates the deep IR physics
it would be interesting to compute fermionic Green functions \cite{Faulkner:2009wj,Cubrovic:2009ye,Alishahiha:2012nm}.
For instance, we could explicitly check the presence of Fermi pockets in the nematic phase within this setup. The dynamics of the universal sector related to these fermions will be given by a couple of Dirac fermions in the bulk with the correct charges with respect to the $U(2)$ gauge symmetry. Finally, constructions from string theory should give further control of the dual field theory we are dealing with \cite{Davis:2011gi, Grignani:2014vaa,Ammon:2009fe,Fadafan:2020fod}.

\section*{Acknowledgments }

We would like to thank Jere Aguilera-Damia, Daniel Areán Byron, Matteo Baggioli, Karl Landsteiner, Yan Liu and Ya-Wen Sun for correspondence. This work was supported in part by the Swedish Research Council, Grant No. VR 2019-04735 (V.J.) and  Fondecyt (Chile) Grant No. 1200399 (R.S.-G.). This work has been funded by the CONICET grants PIP-2017-1109 and PUE 084 ``B\'usqueda de Nueva F\'isica”, and UNLP grants PID-X791 (I.S.L. and N.E.G.).

\bibliographystyle{JHEP}
\bibliography{references}

\end{document}